\documentclass[conference, final, letterpaper]{IEEEtran}

\usepackage{bm}
\usepackage[cmex10]{amsmath}
\usepackage{amssymb}
\usepackage{nicefrac}
\usepackage[space,compress,sort]{cite}
\usepackage{mathtools}
\usepackage{pgf}
\usepackage{tikz}
\usetikzlibrary{matrix}
\usepackage{pgfplots}

\newcommand{\ie}{i.e.,\ }
\newcommand{\vs}{vs.\ }
\newcommand{\defeq}{\ensuremath{\coloneqq}}
\newcommand{\eqdef}{\ensuremath{\eqqcolon}}
\renewcommand{\vec}[1]{\ensuremath{\boldsymbol{#1}}}
\newcommand{\norm}[1]{\lVert#1\rVert}
\newcommand{\R}{\ensuremath{\mathbb{R}}}
\newcommand{\Lo}{\ensuremath{{\mathrm{L1}}}}
\newcommand{\Loc}{\ensuremath{{\Lo_\mathrm{c}}}}
\newcommand{\RS}{\ensuremath{{\mathrm{SI}}}}
\newcommand{\Pd}[1]{\ensuremath{{P_{\mathrm{d}_#1}}}}
\newcommand{\SNR}{\ensuremath{{\mathrm{SNR}}}}
\newcommand{\nSNR}{\ensuremath{{\underbar{\SNR}}}}

\pgfplotsset{compat=1.3}

\begin{document}

\title{Some Comments on the Strong Simplex Conjecture}

\author{
\IEEEauthorblockN{Dejan E. Lazich, Christian Senger, Martin Bossert}
\IEEEauthorblockA{\small Institute of Communications Engineering, Ulm University, Ulm, Germany\\
\{dejan.lazich$\;\vert\;$christian.senger$\;\vert\;$martin.bossert\}@uni-ulm.de}
}

\maketitle

\begin{abstract}
In the disproof of the Strong Simplex Conjecture presented in
\cite{steiner:1994}, a counterexample signal set was found that has higher
average probability of correct optimal decoding than the corresponding regular
simplex signal set, when compared at small values of the signal-to-noise ratio.
The latter was defined as the quotient of average signal \emph{energy} and
average noise \emph{power}. In this paper, it is shown that this interpretation
of the signal-to-noise ratio is inappropriate for a comparison of signal sets,
since it leads to a contradiction with the Channel Coding Theorem. A modified
counterexample signal set is proposed and examined using the classical
interpretation of the signal-to-noise ratio, \ie as the quotient of average signal
energy and average noise energy. This signal set outperforms the regular
simplex signal set for small signal-to-noise ratios without contradicting the
Channel Coding Theorem, hence the Strong Simplex Conjecture remains proven
false.
\end{abstract}

\section{Introduction}\label{sec:intro}

The \emph{Simplex Conjecture (SC)}, one of the oldest and most famous problems of information theory \cite{cover_gopinath:1987}, can be formulated as follows.

Prove that the regular simplex signal set \RS, whose signal vectors are the
$M$ vertices of a regular $N$-dimensional simplex ($M= N+1$) centered at the
origin, is optimal (over all signal sets with $M$ signal vectors) for the
time-discrete \emph{Additive White Gaussian Noise (AWGN)} channel, if
equiprobable signal vectors are used and if the sum over all signal vector
energies is constant.

An optimal signal set maximizes the average probability of correct signal
vector decoding assuming that an optimal decoder is used. The corresponding
optimization constraint is expressed by a constant signal-to-noise ratio. This
ratio is a function of the signal and noise parameters; it should be defined
such that equal transmission conditions for all signal sets under comparison
are guaranteed.

The interest into the SC with its turbulent history\footnote{The history
started with Shannon's comment presented by Rice \cite[p. 68]{rice:1950}, that
despite the fact that a signal set maximizing the smallest distance between
signal vectors (like a regular simplex) leads to a good code, it might not be
the optimal one. See \cite{tanner:1971} for a good overview of the events until
1971 and \cite{steiner:1994} for the events until 1994.} culminated after the
seminal Shannon Lecture ``Towards a proof of the simplex conjecture?'' by
Massey \cite{massey:1988}, presented at the 1988 IEEE International Symposium
on Information Theory in Kobe, Japan. At this occasion, Massey separated the SC
into two variants with different signal energy constraints: The classical
\emph{Weak Simplex Conjecture (WSC)}, where the energies of all signal vectors
are equal (equal-energy-constraint), and the \emph{Strong Simplex Conjecture
(SSC)}, where they are constrained only by an average energy limitation. Before
this, predominantly the WSC was considered in literature. Massey's spark of
interest in the SSC increased the latitude to attack this hard problem.

Indeed, five years after Massey's Shannon Lecture, Steiner proved that ``The strong simplex conjecture is false'' \cite{steiner:1994}. For his disproof, he found a one-dimensional counterexample signal set that outperforms the \RS{} signal set for small values of the signal-to-noise ratio. The validity of the SSC implies the validity of the WSC. However, the invalidity of the SSC does not make any statement about the validity or invalidity of the WSC. Despite this fact, the interest of the information theory community into the classical --- still unsolved --- WSC diminished after Steiner's result.

At the 2nd Asian-European Workshop on Information Theory in 2002, where tribute
was paid to Massey, the first author of the present paper reported about the
progress in solving the WSC and SSC during the period after Massey's Shannon
Lecture \cite{lazic:2002}. At this occasion, a potential inconsistency of the
optimization constraint was revealed. In order to bring clarity into this
possible inconsistency, we recently rechecked and discussed Steiner's results
from \cite{steiner:1994}. Although all his proofs are mathematically correct,
we concluded that slightly modifying the optimization constraint of the SSC
(\ie the interpretation of the signal-to-noise ratio) can cause
ambiguity in the interpretation of Steiner's results. Our conclusion was that
he considered his counterexample signal set under an inadequate optimization
constraint expressed by an interpretation of the signal-to-noise ratio that
penalizes the \RS{} signal set. Thus, his interpretation does not provide equal
transmission conditions for all signal sets under comparison.

In the following section, the main steps of Steiner's disproof are presented
and confirmed by numerical results. In Section~\ref{sec:classical}, we show
that the disproof is not valid any more if the classical definition of the
signal-to-noise ratio is applied. This classical definition was used by Shannon
for the asymptotic comparison of optimal codes in the time-discrete AWGN
channel \cite{shannon:1959}. In Section~\ref{sec:comparison}, we explain why
this fundamental interpretation of the signal-to-noise ratio is appropriate for
a correct examination of the SSC while the interpretation from
\cite{steiner:1994} is not. In Section~\ref{sec:actual}, we introduce a new
counterexample signal set that is a modification of Steiner's and
that actually outperforms the \RS{} signal set for small values of the the
signal-to-noise ratio expressed by Shannon's fundamental interpretation.

\section{Steiner's Disproof of the SSC}\label{sec:steiner}

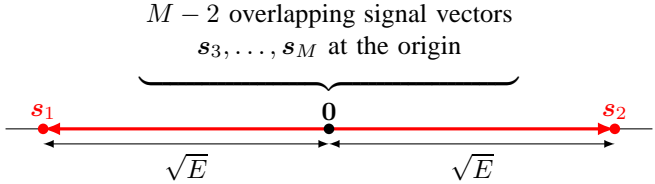
\begin{figure}[htbp]
  \centering
  \begin{tikzpicture}
    
    \def\VsqrtE{3.8}
    
    \draw[] (-\VsqrtE-0.5, 0) -- (\VsqrtE+0.5, 0);
    
    \fill[red] (-\VsqrtE, 0) circle (2pt) node[anchor=south] {$\vec{s}_1$};
    \fill[red] (\VsqrtE, 0) circle (2pt) node[anchor=south] {$\vec{s}_2$};
    \draw[latex-, very thick, draw=red, fill=red] (-\VsqrtE, 0) -- (0, 0);
    \draw[-latex, very thick, draw=red, fill=red] (0, 0) -- (\VsqrtE, 0);
    \fill (0, 0) circle (2pt) node[anchor=south] {$\vec{0}$};

    \draw[latex-latex] (-\VsqrtE, -0.2) -- node[anchor=north] {$\sqrt{E}$} (0, -0.2);
    \draw[latex-latex] (0, -0.2) -- node[anchor=north] {$\sqrt{E}$} (\VsqrtE, -0.2);

    \draw (0, 0.8) node[align=center, anchor=south,below delimiter=\}] {$M-2$ overlapping signal vectors\\$\vec{s}_3, \ldots, \vec{s}_M$ at the origin};

  \end{tikzpicture}
  \caption{Steiner's one-dimensional \Lo{} signal set with $M$ equiprobable signals $\vec{s}_1, \vec{s}_2, \vec{s}_3 \ldots, \vec{s}_M$.}
  \label{fig:L1}
\end{figure}

In \cite{steiner:1994}, Steiner introduced the unusual signal set shown in
Fig.~\ref{fig:L1} and denoted it as \Lo. He showed that it can outperform the
regular simplex signal set \RS{} under a particular optimization constraint as
discussed in the following.

\Lo{} is one-dimensional signal set
consisting of two antipodal signal vectors $\vec{s}_1$ and $\vec{s}_2=
-\vec{s}_1$ having equal energy $E\defeq \norm{\vec{s}_1}^2=\norm{\vec{s}_2}^2$
and $M-2$ additional signal vectors $\vec{s}_3, \ldots, \vec{s}_M$ placed at
the origin\footnote{\label{fn:critics}Some critics of Steiner's disproof claim
that the $M-2$ overlapping signal vectors must be regarded as a single signal
vector with higher a-priori probability equal to $\nicefrac{(M-2)}{M}$. Using
this point of view, it is easy to show that \Lo{} never performs better than
the \RS{} signal set under Steiner's optimization constraint. However, this
result does not invalidate Steiner's disproof because a minor variation of the
\Lo{} signal set, where the signal vectors $\vec{s}_3, \ldots, \vec{s}_M$ are
displaced from the origin by an arbitrarily small $\varepsilon>0$, renders them
again equiprobable and mutually distinct.}, such that
$\norm{\vec{s}_3}^2=\cdots=\norm{\vec{s}_M}^2=0$. It is assumed that the signal
vectors from \Lo{} are i.i.d., having equal a-priori probabilities
$\Pr(\vec{s}_m)=\nicefrac{1}{M}$, $m=1, \ldots, M$, so that in this case the
minimum distance decoder is optimal and the average (expected) signal energy of
the \Lo{} signal set is
\begin{equation}\label{eqn:deflambdasquare}
	\mathbb{E}\left\{\norm{\vec{s}_m}^2\right\}_{m=1}^M=\frac{2E}{M}\eqdef\lambda^2.
\end{equation}
The denotation $\lambda^2$ for the average signal energy of a
signal set is adopted from \cite{steiner:1994}. In particular, Steiner used
$\nicefrac{\lambda^2}{\sigma^2}$ under the assumption $\sigma^2= 1$ as
signal-to-noise ratio and thus optimized subject to $\lambda^2=\mathrm{const}$,
where $\sigma^2$ is the variance of the time-discrete AWGN.

The average probability $\Pd{\Lo}$ of correct decoding for an optimally decoded \Lo{} signal set with $M$ equiprobable signal vectors is given by \cite[Eqn.~(15)]{steiner:1994}, \ie
\begin{equation}\label{eqn:PdSteinerL1}
  \Pd{\Lo}\defeq\frac{1}{M}%
    \left[%
      4\Phi%
        \left(\sqrt{\frac{\lambda^2 M}{8}}\right)-1%
    \right],\quad M\geq 3,
\end{equation}
where 
\begin{equation*}
  \Phi(x)\defeq \frac{1}{\sqrt{2\pi}}%
    \int\limits_{-\infty}^{x}\exp\left(-\frac{u^2}{2}\right)\mathrm{d}u
\end{equation*}
is the cumulative distribution function of the time-discrete AWGN with variance $\sigma^2= 1$ and zero mean.

The average probability of correct decoding for an optimally decoded \RS{} signal set with $M= N+1$ equiprobable signal vectors is given by \cite[Eqn.~(10)]{steiner:1994}, \ie
\begin{multline}\label{eqn:PdSteinerRS}
  \Pd{\RS}\hspace{-0.1cm}\defeq\hspace{-0.1cm} \frac{1}{\sqrt{2\pi}}%
    \int\limits_{-\infty}^{\infty}\hspace{-0.2cm}\exp%
      \left[%
        -\frac{\left(x-\sqrt{\lambda^2 \frac{M}{M-1}}\right)^2}{2}
      \right]%
      \hspace{-0.15cm}\left[
        \Phi(x)
      \right]^{M-1}%
    \mathrm{d}x,\\M\geq 2,
\end{multline}
where $\lambda^2$ represents the energy of each signal vector, and thus also the average energy of the signals in the \RS{} signal set. Eqn.~(\ref{eqn:PdSteinerRS}) originates from Weber, who completely derived it in \cite[Eqn.~(14.31)]{weber:1987}.

Steiner's major result, the counterexample \Lo{} signal set for the SSC, was presented in \cite[Section~III]{steiner:1994}. Using analytical methods and some numerical evaluations, he showed that the \Lo{} signal set can outperform the \RS{} signal set for all $M\geq 7$ under the average energy optimization constraint $\lambda^2=\mathrm{const}$, which corresponds to his interpretation of the signal-to-noise ratio, \ie $\nicefrac{\lambda^2}{\sigma^2}$ with $\sigma^2= 1$. A comparison of (\ref{eqn:PdSteinerL1}) and (\ref{eqn:PdSteinerRS}) showed that they are guaranteed to have a crossing point $\lambda^2_\mathrm{X}(M)$. Thus, for average signal set energies $\lambda^2$ in the interval $0\leq\lambda^2<\lambda^2_\mathrm{X}(M)$ and $M\geq 7$ signal vectors, \Lo{} outperforms \RS{} in terms of the average probability of correct optimal decoding.

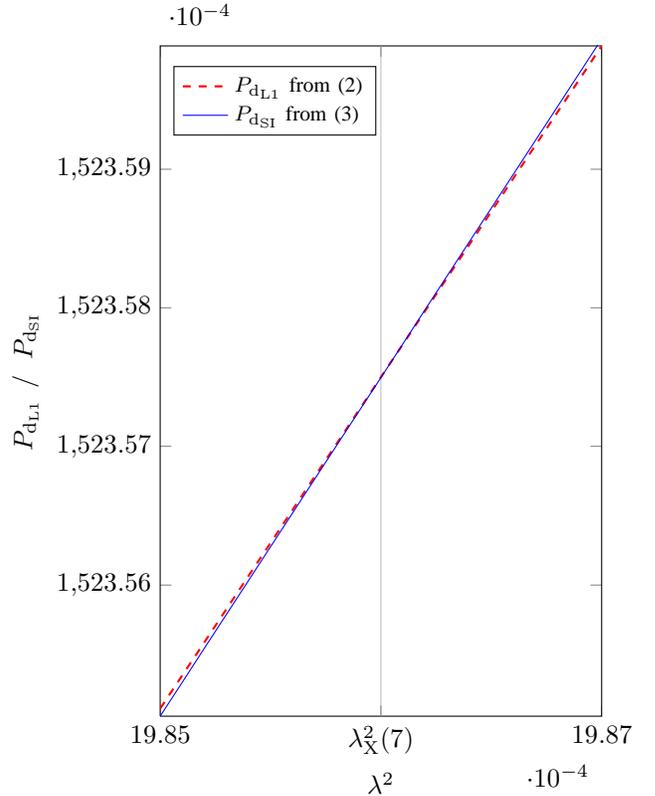
\begin{figure}[htbp]
  \centering
  \begin{tikzpicture}

    \begin{axis}[
    xlabel=$\lambda^2$,
    xtick={0.0019850, 0.0019870},
    ylabel=$\Pd{\Lo} \;\left/\right.\; \Pd{\RS}$,
    ytick={0.152356, 0.152357, 0.152358, 0.152359, 0.152360},
    legend pos=north west,
    width=212pt,
    height=10.5cm,
    enlarge x limits=false,
    enlarge y limits=false,
    extra x ticks={0.001986},
    extra x tick labels={$\lambda^2_\mathrm{X}(7)$},
    extra x tick style={grid=major},
    scaled ticks=base 10:4,
    legend style={cells={anchor=west}}
    ]

    \addplot+[draw=red,no marks,dashed,thick] file {./mathematica/senger_PdSteinerL1.dat};
    \addlegendentry{\footnotesize $\Pd{\Lo}$ from (\ref{eqn:PdSteinerL1})};

    \addplot+[draw=blue,no marks,solid] file {./mathematica/senger_PdSteinerRS.dat};
    \addlegendentry{\footnotesize $\Pd{\RS}$ from (\ref{eqn:PdSteinerRS})};

    \end{axis}
    
  \end{tikzpicture}
  \caption{Average probability of correct decoding for $M= 7$ equiprobable signal vectors from signal sets \Lo{} and \RS{} \vs the average signal set energy $\lambda^2$. Both signal sets are decoded using optimal minimum distance decoders.}
  \label{fig:comparisonSteiner}
\end{figure}

Indeed, the probability curves for correct decoding of the \Lo{} and \RS{} signal sets for $M= 7$ in Fig.~\ref{fig:comparisonSteiner} clearly show a crossing point at $\lambda^2_\mathrm{X}(7)\approx 19.86\cdot 10^{-4}$. 

In further numerical evaluations, we could not find crossing points for $3\leq M<7$, while crossing points for $M\geq 7$ were always found, \ie \Lo{} performed better than \RS{} for $\lambda^2$ in the interval $0\leq\lambda^2<\lambda^2_\mathrm{X}(M)$. In all evaluated cases, our numerical results coincide with Steiner's analytical result if his interpretation of the signal-to-noise ratio ($\nicefrac{\lambda^2}{\sigma^2}$ with $\sigma^2= 1$) is used in the optimization constraint.

However, several interpretations of the signal-to-noise ratio are used in literature, sometimes causing confusion and a lack of comparability of results. In the following section, we analyze Steiner's results and the SSC using Shannon's original interpretation of the signal-to-noise ratio.

\section{Average Probability of Correct Decoding using the Classical Signal-to-Noise Ratio}\label{sec:classical}

Shannon derived upper and lower bounds on the reliability function (error
exponent) of the time-discrete AWGN channel using the classical signal-to-noise
ratio interpretation \cite{shannon:1959}\footnote{Shannon called $\sigma^2$
\emph{noise power} and denoted it by $\mathrm{N}$.}
\begin{equation*}
  A^2\defeq \frac{P}{\sigma^2}.
\end{equation*}
He called $P$ the \emph{signal power} and assumed that each signal vector is on the surface of a sphere of radius $\sqrt{NP}$. Consequently, $NP=\norm{\vec{s}_m}^2$, $m=1, \ldots, M$, represents the energy of equal-energy signal vectors from a considered signal set (called code in \cite{shannon:1959}). The average energy of noise vectors of length $N$ produced by the time-discrete AWGN channel is $N\sigma^2$, such that
\begin{equation*}
  A^2=\frac{NP}{N\sigma^2}=\frac{P}{\sigma^2}.
\end{equation*}

In \cite{shannon:1959}, the capacity $C$ of the time-discrete AWGN channel (per degree of freedom, \ie dimension) was expressed by
\begin{equation*}
  C=\frac{1}{2}\log\left(A^2+1\right).
\end{equation*}
Consequently, $A^2$ represents the fundamental interpretation of the
signal-to-noise ratio if equiprobable and equal-energy signal
sets\footnote{Note that the capacity of the time-discrete AWGN channel is the
same, no matter if the signal vectors are average- or equal-energy
constrained \cite{shannon:1959}.} are used, since it is involved in the Channel
Coding Theorem and the capacity of the time-discrete AWGN channel.

In the general case, the definition of this fundamental signal-to-noise ratio (which we denote as \SNR) for the time-discrete AWGN channel (and similar channel models) is given by
\begin{equation}\label{eqn:SNR}
  \SNR\defeq\frac{\sum_{m=1}^{M} \Pr(\vec{s}_m)\norm{\vec{s}_m}^2}{N\sigma^2},
\end{equation}
where $\Pr(\vec{s}_m)$ is the a-priori probability, $\norm{\vec{s}_m}^2$ is the energy of the signal vector $\vec{s}_m$, $m=1, \ldots, M$, and $\sigma^2$ is the variance of a zero-mean Gaussian random variable $n_i$, $i=1, \ldots, N$ that is one of the $N$ components of the time-discrete AWGN vector $\vec{n}=(n_1, \ldots, n_N)$.

For equiprobable signal vectors, (\ref{eqn:SNR}) reduces to
\begin{equation}\label{eqn:equiprobableSNR}
  \SNR=\frac{\sum_{m=1}^M\norm{\vec{s}_m}^2}{MN\sigma^2},
\end{equation}
so that the fundamental \SNR{} for the one-dimensional \Lo{} signal set ($N=1$) becomes
\begin{equation*}
  \SNR=\frac{2E}{M\sigma^2}=\frac{\lambda^2}{\sigma^2},
\end{equation*}
where the last equality follows from (\ref{eqn:deflambdasquare}). By setting $\sigma^2= 1$ as in \cite{steiner:1994}, the \emph{normalized fundamental \SNR{}} (denoted as \nSNR) for the \Lo{} signal set becomes
\begin{equation*}
  \nSNR=\lambda^2.
\end{equation*}
Thus, $\nSNR$ in the one-dimensional case reduces to the interpretation of the signal-to-noise ratio used in \cite{steiner:1994} by Steiner.

However, the normalized classical \SNR{} for the $N$\hbox{-}dimensional \RS{} signal set that consists of $M=N+1$ equiprobable signal vectors becomes
\begin{equation}\label{eqn:RSnormSNR}
  \nSNR=\frac{\lambda^2}{N},
\end{equation}
which is obtained by inserting $\norm{\vec{s}_1}^2=\cdots=\norm{\vec{s}_M}^2=\lambda^2$  into (\ref{eqn:equiprobableSNR}) and then setting $\sigma^2= 1$. By inserting (\ref{eqn:RSnormSNR}) into (\ref{eqn:PdSteinerRS}), we obtain the average probability
\begin{multline}\label{eqn:PdclassicalRS}
  \Pd{\RS}\hspace{-0.1cm}\defeq\hspace{-0.1cm}\frac{1}{\sqrt{2\pi}}\int\limits_{-\infty}^{\infty}%
    \hspace{-0.2cm}\exp%
    \left[%
      -\frac{%
        \left(%
          x-\sqrt{M\cdot\nSNR}%
        \right)^2}{2}%
    \right]%
    \hspace{-0.15cm}\left[
      \Phi(x)
    \right]^{M-1}%
    \mathrm{d}x,\\M\geq 2
\end{multline}
of correct decoding for an optimally decoded \RS{} signal set with equiprobable signal vectors.

\begin{figure}[htbp]
  \centering
  \begin{tikzpicture}

    \begin{axis}[
    xlabel=\nSNR{},
    xmin=0.0,
    xmax=2.0,
    ymin=0.0,
    ymax=1.0,
    ylabel=$\Pd{\Lo} \;\left/\right.\; \Pd{\RS}$,
    legend pos=south east,
    width=212pt,
    height=7cm,
    enlarge x limits=false,
    enlarge y limits=false,
    grid=major,
    legend style={cells={anchor=west}}
    ]

    \addplot+[draw=red,no marks,dashed] file {./mathematica/senger_PdclassicalL1_M7.dat};
    \addlegendentry{\footnotesize $\Pd{\Lo}$ from (\ref{eqn:PdSteinerL1})};

    \addplot+[draw=blue,no marks,solid] file {./mathematica/senger_PdclassicalRS_M7.dat};
    \addlegendentry{\footnotesize $\Pd{\RS}$ from (\ref{eqn:PdclassicalRS})};
    
    \end{axis}

  \end{tikzpicture}
  \caption{Average probability of correct decoding for $M= 7$ equiprobable signal vectors from signal sets \Lo{} and \RS{} \vs \nSNR. Both signal sets are decoded using optimal minimum distance decoders.}
  \label{fig:comparisonclassical}
\end{figure}
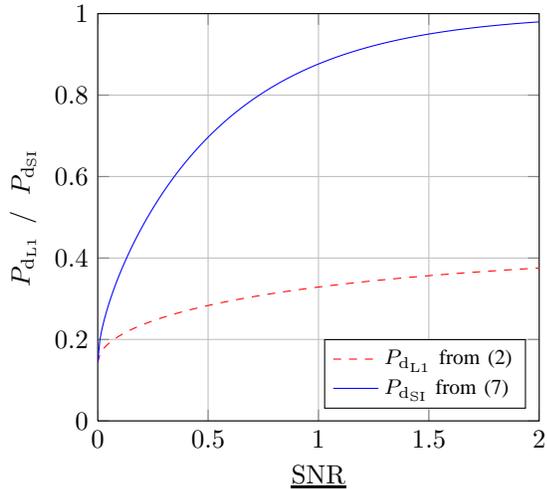

Consequently, if we aim to compare signal sets \Lo{} and \RS{} according to \nSNR{}\footnote{This fundamental interpretation of the signal-to-noise ratio was used by Shannon for the asymptotical comparison ($N\rightarrow\infty$) of optimal codes in the time-discrete AWGN channel in \cite{shannon:1959}.}, we have to compare (\ref{eqn:PdSteinerL1}) to (\ref{eqn:PdclassicalRS}) instead of (\ref{eqn:PdSteinerL1}) to (\ref{eqn:PdSteinerRS}). The probability curves of correct decoding for $M= 7$ and signal sets \Lo{} (Eqn.~(\ref{eqn:PdSteinerL1}) with $\lambda^2=\nSNR$) and \RS{} (Eqn.~(\ref{eqn:PdclassicalRS}) with $\nSNR=\nicefrac{\lambda^2}{N}$, $N= M-1$) \vs \nSNR{} are shown in Fig.~\ref{fig:comparisonclassical}.

It can be seen at first glance that now the \RS{} signal set outperforms the \Lo{} signal set for all values of \nSNR. Furthermore, in spite of intensive search, no crossing point (corresponding to $\lambda^2_\mathrm{X}(7)$ in Fig.~\ref{fig:comparisonSteiner}) could be found for $M= 7$. Likewise, we could not find an $M\geq 3$ for which the \Lo{} signal set outperforms the \RS{} signal set for any \nSNR.

A complete proof of this observation requires analytical methods. However, already the numerical results at hand allow to conclude that the disproof of the SSC in \cite{steiner:1994} is not valid if the fundamental interpretation of the signal-to-noise ratio is used.

\section{Which Interpretation of the Signal-to-Noise Ratio is Appropriate for an Examination of the SSC?}\label{sec:comparison}

Our observations in the previous section obviously contradict Steiner's disproof of the SSC from \cite{steiner:1994} if the fundamental interpretation of the signal-to-noise ratio is used in the optimization constraint. One could ask which interpretation is the correct one for an examination of the SSC, Steiner's $\lambda^2$ or the normalized fundamental $\nSNR=\nicefrac{\lambda^2}{N}$?

\begin{figure}[htbp]
  \centering
  \begin{tikzpicture}
    
    \begin{axis}[
    xlabel=\nSNR{},
    xmin=0.0,
    xmax=6.0,
    ymin=0.0,
    ymax=1.0,
    ylabel=$\Pd{\RS}$,
    legend pos=south east,
    width=212pt,
    height=7cm,
    enlarge x limits=false,
    enlarge y limits=false,
    grid=major,
    legend style={cells={anchor=west}}
    ]

    \addplot+[draw=blue,no marks,solid] file {./mathematica/senger_PdclassicalRS_M3.dat};
    \node[align=left,text width=2cm,pin={170, pin distance=0.34cm:{}},inner sep=2pt] at (axis cs:3.3, 0.85) {$M= 3$};

    \addplot+[draw=green,no marks,solid] file {./mathematica/senger_PdclassicalRS_M7.dat};
    \node[align=left,text width=2cm,pin={170, pin distance=1.31cm:{}},inner sep=2pt] at (axis cs:3.3, 0.75) {$M= 7$};

    \addplot+[draw=red, no marks,solid] file {./mathematica/senger_PdclassicalRS_M20.dat};
    \node[align=left,text width=2cm,pin={170, pin distance=1.82cm:{}},inner sep=2pt] at (axis cs:3.3, 0.65) {$M=20$};

    \addplot+[draw=black,no marks,solid] file {./mathematica/senger_PdclassicalRS_M30.dat};
    \node[align=left,text width=2cm,pin={170, pin distance=1.98cm:{}},inner sep=2pt] at (axis cs:3.3, 0.55) {$M= 30$};

    \end{axis}

  \end{tikzpicture}
  \caption{Average probability of correct decoding for $M= 3, 7, 20, 30$ equiprobable signal vectors with corresponding dimensions $N= 2, 6, 19, 29$ from the \RS{} signal sets \vs \nSNR. All signal sets are decoded using optimal minimum distance decoders.}
  \label{fig:comparisonclassicalM}
\end{figure}
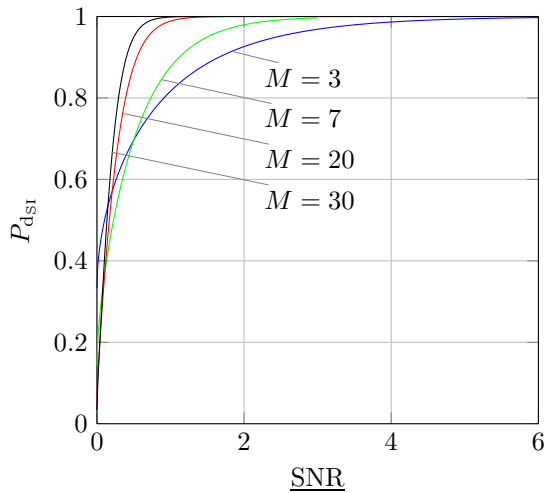

One way out of this dilemma is to evaluate the average probability of correct decoding for the \RS{} signal set for different values of $M$. Fig.~\ref{fig:comparisonclassicalM} shows the corresponding $\Pd{\RS}$ \vs $\nSNR$; the curves were calculated using (\ref{eqn:PdclassicalRS}). Clearly, larger dimension $N= M-1$ leads to larger average probability of correct optimal decoding at sufficiently large \nSNR. This is in line with the Channel Coding Theorem \cite{shannon:1959} and Ziv's result \cite{ziv:1964} that equal-energy signal sets, which --- like the \RS{} signal set --- maximize the smallest Euclidean distance between signal vectors are optimal if \nSNR{} is large enough.

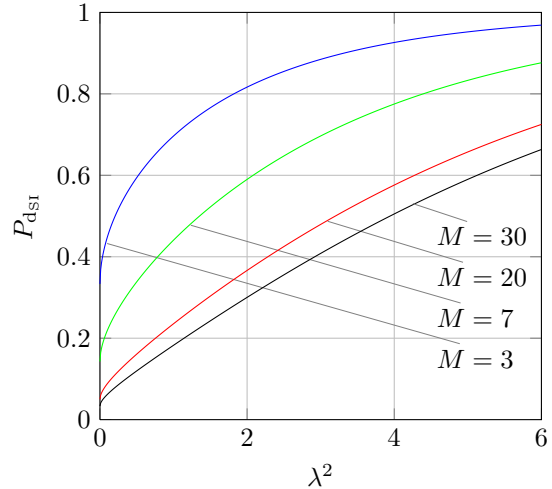
\begin{figure}[htbp]
  \centering
  \begin{tikzpicture}

    \begin{axis}[
    xlabel=$\lambda^2$,
    xmin=0.0,
    xmax=6.0,
    ymin=0.0,
    ymax=1.0,
    ylabel=$\Pd{\RS}$,
    legend pos=north west,
    width=212pt,
    height=7cm,
    enlarge x limits=false,
    enlarge y limits=false,
    grid=major,
    legend style={cells={anchor=west}}
    ]

    \addplot+[draw=blue,no marks,solid] file {./mathematica/senger_PdSteinerRS_M3.dat};
    \node[align=left,text width=2cm,pin={165, pin distance=4.8cm:{}},inner sep=2pt] at (axis cs:5.6, 0.15) {$M= 3$};

    \addplot+[draw=green,no marks,solid] file {./mathematica/senger_PdSteinerRS_M7.dat};
    \node[align=left,text width=2cm,pin={165, pin distance=3.65cm:{}},inner sep=2pt] at (axis cs:5.6, 0.25) {$M= 7$};

    \addplot+[draw=red, no marks,solid] file {./mathematica/senger_PdSteinerRS_M20.dat};
    \node[align=left,text width=2cm,pin={165, pin distance=1.77cm:{}},inner sep=2pt] at (axis cs:5.6, 0.35) {$M= 20$};

    \addplot+[draw=black,no marks,solid] file {./mathematica/senger_PdSteinerRS_M30.dat};
    \node[align=left,text width=2cm,pin={165, pin distance=0.6cm:{}},inner sep=2pt] at (axis cs:5.6, 0.45) {$M= 30$};

    \end{axis}

  \end{tikzpicture}
  \caption{Average probability of correct decoding for $M= 3, 7, 20, 30$ equiprobable signal vectors with corresponding dimensionality $N= 2, 6, 19, 29$ from the \RS{} signal sets \vs signal-to-noise ratio $\lambda^2$ as used by Steiner \cite{steiner:1994}. All signal sets are decoded using optimal minimum distance decoders.}
  \label{fig:comparisonSteinerlM}
\end{figure}

The situation is different if we consider the average probability of correct decoding $\Pd{\RS}$ \vs $\lambda^2$ using (\ref{eqn:PdSteinerRS}). The corresponding curves in Fig.~\ref{fig:comparisonSteinerlM} show that increasing the dimension $N$ of the \RS{} signal set actually leads to decreasing probability of correct optimal decoding for all values of $\lambda^2$. This is in contradiction to the fundamental results of Shannon \cite{shannon:1959} and Ziv \cite{ziv:1964}. From this contradiction, we conclude that Steiner's interpretation of the signal-to-noise ratio is not appropriate for an examination of the SSC.

The consequence of this observation alone would be that the SSC is still an open problem.

\section{A Signal Set that Actually Disproves the SSC}\label{sec:actual}

Besides the inappropriate interpretation of the signal-to-noise ratio as
described in the previous section, another deficiency of the disproof in
\cite{steiner:1994} is evident. Fortunately, the correction of this second flaw
allows us to rectify Steiner's claim that the SSC is false using a different
counterexample signal set. In doing so, we convert two failures into a success. 

\begin{figure*}[t]
  \centering
   \begin{tikzpicture}

    \def\VsqrtE{2.8}

    \clip (2, 6.8) rectangle (20, 12.2);

    \begin{axis}[
      xtick=\empty,
      xmin=-2.8,
      xmax=2.8,
      ytick=\empty,
      ymin=-2.8,
      ymax=2.8,
      ztick=\empty,
      zmin=-2.8,
      zmax=2.8,
      view={15}{20},
      axis lines=center,
      width=1.3\textwidth,
      ]

      \addplot3+[mesh,no markers,draw=black!40,domain=-1:1,samples=10,y domain=-0.4:1,samples y=10]
        ( {0.6-0.15*x-0.25*y}, {x}, {y-0.1} );
      \node[] at (axis cs:-0.1, 1.2, 0.5) {$P_2$};
        
      \addplot3+[mesh,no markers,draw=black!40,domain=-1:1,samples=10,y domain=-0.4:1,samples y=10]
        ( {-0.6-0.15*x-0.25*y}, {x}, {y-0.4} );
       \node[] at (axis cs:-0.8, 1, 0.4) {$P_1$};

      \pgfmathparse{0.6*\VsqrtE}
      \pgfmathsetmacro{\Vx}{\pgfmathresult}
      \pgfmathparse{0.15*\VsqrtE}
      \pgfmathsetmacro{\Vy}{\pgfmathresult}
      \pgfmathparse{0.25*\VsqrtE}
      \pgfmathsetmacro{\Vz}{\pgfmathresult}

      \addplot3[mark=*,draw=red,fill=red,solid,very thick,-latex,mark options={fill=red,scale=1.0,solid}] coordinates {%
        (0,0,0)
        (\Vx, \Vy, \Vz)
        };
      \addplot3[mark=*,draw=red,fill=red,solid,very thick,-latex,mark options={fill=red,scale=1.0,solid}] coordinates {%
        (0,0,0)
        (-\Vx, -\Vy, -\Vz)
        };
      \node[pin=0:{\color{red} $\vec{s}'_1$}] at (axis cs:\Vx, \Vy, \Vz) {};
      \node[pin=180:{\color{red} $\vec{s}'_2$}] at (axis cs:-\Vx, -\Vy, -\Vz) {};
      \node[pin=110:{$\vec{s}'_3$}] at (axis cs:0,0,0) {};
      \node[pin=155:{$\vec{s}'_4$}] at (axis cs:0,0,0) {};
      
      \addplot3+[mesh,no markers,thick,dotted,draw=red,domain=0:1,samples=2,y domain=0:1,samples y=2]
        ( {\Vx*x}, {\Vy*y}, {0} );
      \addplot3+[mesh,no markers,thick,dotted,draw=red,domain=0:1,samples=2,y domain=0:1,samples y=2]
        ( {\Vx*x}, {\Vy*y}, {\Vz} );
      \addplot3+[mesh,no markers,thick,dotted,draw=red,domain=0:1,samples=2,y domain=0:1,samples y=2]
        ( {0}, {\Vy*x}, {\Vz*y} );
      \addplot3+[mesh,no markers,thick,dotted,draw=red,domain=0:1,samples=2,y domain=0:1,samples y=2]
        ( {\Vx}, {\Vy*x}, {\Vz*y} );
        
      \addplot3+[mesh,no markers,thick,dotted,draw=red,domain=0:1,samples=2,y domain=0:1,samples y=2]
        ( {-\Vx*x}, {-\Vy*y}, {0} );
      \addplot3+[mesh,no markers,thick,dotted,draw=red,domain=0:1,samples=2,y domain=0:1,samples y=2]
        ( {-\Vx*x}, {-\Vy*y}, {-\Vz} );
      \addplot3+[mesh,no markers,thick,dotted,draw=red,domain=0:1,samples=2,y domain=0:1,samples y=2]
        ( {0}, {-\Vy*x}, {-\Vz*y} );
      \addplot3+[mesh,no markers,thick,dotted,draw=red,domain=0:1,samples=2,y domain=0:1,samples y=2]
        ( {-\Vx}, {-\Vy*x}, {-\Vz*y} );

    \end{axis}
  \end{tikzpicture}
  \caption{An example for the coded \Loc{} signal set in general position of $\R^3$ with $M=4$ signal vectors. It is one-dimensional but every signal vector is transmitted with $N_\mathrm{u}=3$ real-valued channel uses. The two planes $P_1$ and $P_2$ represent the boundaries of the decision regions of optimal decoding.}
  \label{fig:Loc}
\end{figure*}
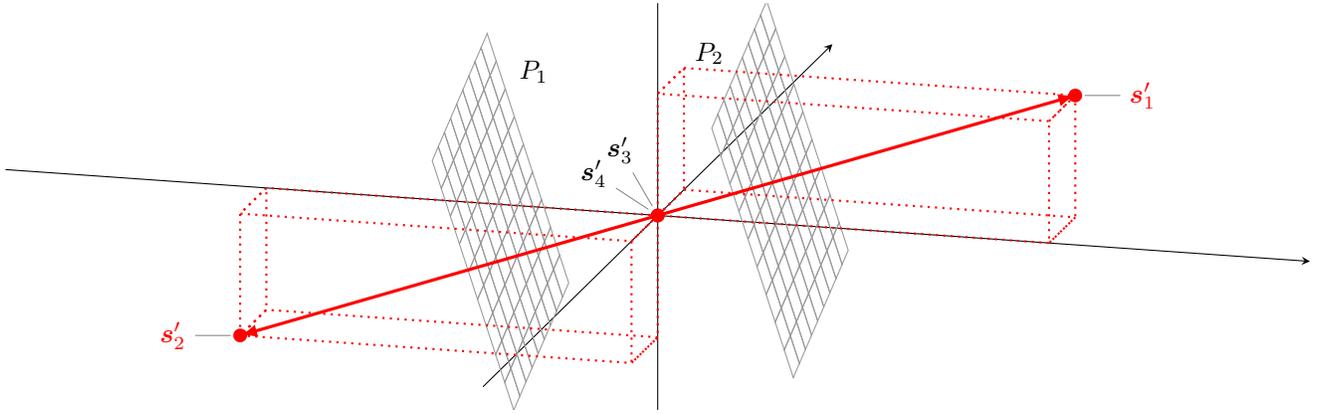

The \Lo{} counterexample signal set is restricted to one single dimension for
all values of $M\geq 7$. Thus, all signal vectors from \Lo{} can be transmitted
by only one real-valued channel use (\ie $-\sqrt{E}$, $0$, or $\sqrt{E}$). At
the same time, signal vectors from the \RS{} signal set must by transmitted by
$N_\mathrm{u}=N=M-1$ real-valued channel uses. Hence, the
code rate (generally defined by $R\defeq\nicefrac{\log_2(M)}{N_\mathrm{u}}$ for
signal sets with $M$ equiprobable signal vectors) of \RS{} is
$R_\RS\defeq\nicefrac{\log_2(M)}{(M-1)}$ while for the \Lo{} signal set it is
$R_\Lo\defeq\log_2(M)$. Obviously, $R_\Lo> R_\RS$ for $M\geq 3$.

Consequently, disproving the SSC using the \Lo{} signal set and the frequent
signal-to-noise ratio interpretation $\nicefrac{E_\mathrm{b}}{N_0}$
\cite{forney:2003} would be inappropriate as well. Since
$\nicefrac{E_\mathrm{b}}{N_0}=\nicefrac{\SNR}{2R}$, where
$N_0\defeq\nicefrac{\sigma^2}{2}$, this interpretation of the signal-to-noise
ratio involves the code rate, thus rendering a comparison between the \Lo{} and
\RS{} signal sets unfair due to the unequal transmission conditions of the signal sets under comparison.

In order to ameliorate this deficiency of the \Lo{} signal set and to
''rescue`` Steiner's elegant counterexample, we introduce a new signal set
\Loc, which we refer to as \emph{coded} \Lo. It is a slight but substantial
modification of \Lo{} that remains its basic structure\footnote{\Loc{} consists
of two equal-energy antipodal signal vectors $\vec{s}'_1=-\vec{s}'_2$ with
signal energy $E\defeq\norm{\vec{s}'_1}^2=\norm{\vec{s}'_2}^2$ and $M-2$ signal
vectors $\vec{s}'_3, \ldots, \vec{s}'_M$ at the origin. All signal vectors are
equiprobable. Critics who claim that the $M-2$ overlapping signal vectors must
be regarded as one single signal vector with higher a-priori probability are
referred to footnote~\ref{fn:critics}. Note that, due to the distribution of
the signal energy over the $N_\mathrm{u}=M-1$ real-valued channel uses, the \Loc{} signal set may
have significantly smaller peak power compared to \Lo.} and is rotated to a
general position in $\R^N$, see Fig.\ref{fig:Loc}. In this way, \Loc{} is still
one-dimensional\footnote{That is, the signal vectors from \Loc{} span a
one-dimensional subspace.} but its signal vectors $\vec{s}'_1$ and $\vec{s}'_2$
(corresponding to the signal vectors $\vec{s}_1$ and $\vec{s}_2$ of \Lo) must
be transmitted with $N_\mathrm{u}=M-1$ real-valued channel uses.  Clearly, this
renders the code rates of \RS{} and \Loc{} equal, \ie
$R_\RS=R_\Loc=\nicefrac{\log_2(M)}{(M-1)}$.

Since the \Loc{} signal set requires $N_\mathrm{u}=M-1$ real-valued channel
uses, the average energy of the AWGN noise vectors that affect the transmitted
signal vectors is $N\sigma^2$, so that the fundamental signal-to-noise ratio
normalized by $\sigma^2= 1$ becomes as in (\ref{eqn:RSnormSNR}).
By inserting (\ref{eqn:RSnormSNR}) in (\ref{eqn:PdSteinerL1}) (which is
allowed, since rotation of a signal set does not change its probability of
correct decoding), we obtain
\begin{equation}\label{eqn:PdcodedL1}
  \Pd{\Loc}\defeq\frac{1}{M}%
    \left[%
      4\Phi%
        \left(\sqrt{\frac{(M-1)M\cdot\nSNR}{8}}\right)-1%
    \right],\quad M\geq 3.
\end{equation}

By evaluating (\ref{eqn:PdclassicalRS}) and (\ref{eqn:PdcodedL1}) for $M=
7$, a crossing point of the probability curves for correct optimal
decoding at $\nSNR_\mathrm{X}(7)\approx 3.3\cdot 10^{-4}$ can be observed. In further numerical
evaluations, we could not find crossing points for $3\leq M<7$, while crossing
points for $M\geq 7$ were always found, \ie \Loc{} performed better than \RS{}
for \nSNR{} in the interval $0\leq\nSNR<\nSNR_\mathrm{X}(M)$.

Our numerical evaluation qualitatively supports Steiner's claim but differ
quantitatively, \ie we obtain different probability curves and crossing points.
Consequently, the Strong Simplex Conjecture is \emph{indeed} false.

\section{Conclusions}\label{sec:conclusions}

We started the paper with a recapitulation of Steiner's disproof of the SSC and
showed that his counterexample \Lo{} signal set cannot outperform the regular
simplex signal set, when the comparison is based on the classical normalized
signal-to-noise ratio \nSNR{}. In order to establish that \nSNR{} is the
appropriate interpretation of the signal-to-noise ratio for an examination of
the SSC, we showed that the interpretation used in \cite{steiner:1994} leads to
a contradiction with the Channel Coding Theorem.  However, we managed to
rectify Steiner's claim of the SSC's invalidity by introducing a slightly but
substantially modified counterexample signal set \Loc, whose $M$ signal vectors
have to be transmitted with $M-1$ real-valued channel uses. We argue that
\Loc{} outperforms the regular simplex signal set for small values of \nSNR{}
whenever $M\geq 7$.

\vspace{0.7cm}


\def\noopsort#1{}

\end{document}